\PassOptionsToPackage{table}{xcolor}
\documentclass[sigconf, nonacm]{acmart}
\AtBeginDocument{%
  }

\usepackage{multirow}
\usepackage{tabularx}

\begin{document}

\title{TriGen: A Generative Paradigm for Trigger-Induced Recommendation}

\author{Muyu Zou}
\authornote{These authors contributed equally to this research.}
\affiliation{%
  \institution{Alibaba International Digital Commerce Group}
  \country{China}
}
\email{zoumuyu.zmy@alibaba-inc.com}

\author{Haibo Xing}
\authornotemark[1]
\affiliation{%
  \institution{Alibaba International Digital Commerce Group}
  \city{Hangzhou}
  \country{China}
}
\email{xinghaibo.xhb@alibaba-inc.com}

\author{Shunye Tang}
\authornotemark[1]
\affiliation{%
  \institution{Nankai University}
  \department{College of Software}
  \city{Tianjin}
  \country{China}
}
\email{2113298@mail.nankai.edu.cn}

\author{Liyan Tie}
\affiliation{%
  \institution{Independent Researcher}
  \country{}
}

\author{Lingyu Mu}
\affiliation{%
  \institution{Alibaba International Digital Commerce Group}
  \city{Beijing}
  \country{China}
}
\email{moulingyu.mly@alibaba-inc.com}

\author{Yandong Wang}
\affiliation{%
  \institution{Alibaba International Digital Commerce Group}
  \country{China}
}
\email{wangyandong.wyd@alibaba-inc.com}

\author{Jinxin Hu}
\authornote{Corresponding author.}
\affiliation{%
  \institution{Alibaba International Digital Commerce Group}
  \city{Beijing}
  \country{China}
}
\email{jinxin.hjx@lazada.com}

\renewcommand{\shortauthors}{Muyu Zou et al.}

\begin{abstract}
Generative recommendation typically operates under the User-Induced Recommendation (UIR) setting, predicting target items by autoregressively generating semantic IDs as a passive continuation of long-term user history. However, this paradigm overlooks Trigger-Induced Recommendation (TIR), an essential industrial scenario where a user click triggers an immediate follow-up feed. In TIR, recommendations must dynamically align with the instant trigger intent without compromising historical preference continuity. Traditional TIR methods rely on discriminative scoring and cannot guide token-level generation, while existing generative models lack mechanisms to reconcile this structural tension. To address this challenge, we propose TriGen, establishing a tailored generative paradigm for trigger-induced recommendation. TriGen introduces calibrated tail injection to re-encode local history, adaptively arbitrates between persistent tastes and immediate impulses via dynamic gating, and applies position-specific transformations to sustain level-aware trigger guidance across decoding steps. Extensive offline evaluations confirm consistent gains over competitive baselines. Furthermore, online A/B tests across six country markets demonstrate significant revenue and GMV improvements, leading to the full production deployment of TriGen. 
\end{abstract}



\maketitle

\section{Introduction}

\begin{figure}
  \centering
  \includegraphics[width=\columnwidth]{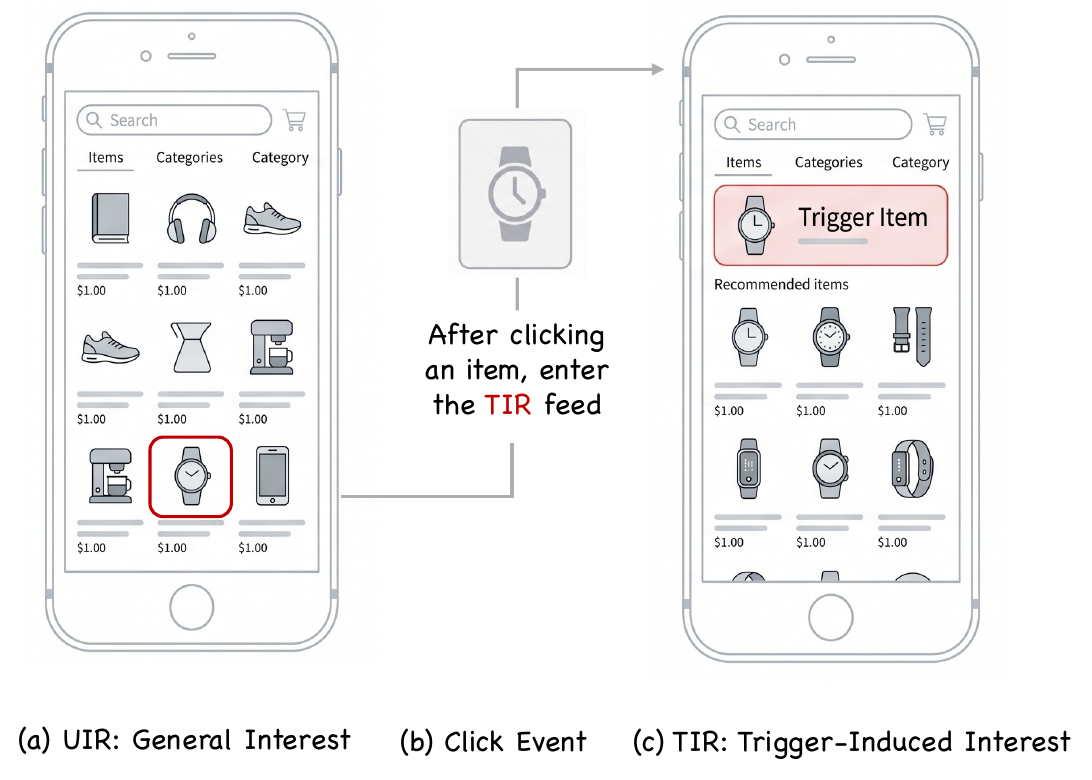}
  \caption{Trigger-Induced Recommendation (TIR) in an e-commerce
    application. The item the user clicks (b) in the general feed (a) becomes
    the trigger of the follow-up feed (c), where it is pinned above the
    recommendations organized around it.}
  \Description{Two mobile phone screens connected by an arrow. The left
    screen, labelled UIR general interest, shows a feed of items from mixed
    categories such as a book, headphones, shoes, a coffee maker and a phone,
    with a watch outlined in red as the item the user clicks. Between the two
    screens the clicked watch appears again, next to a caption saying that
    clicking an item enters the TIR feed. The right screen, labelled TIR
    trigger-induced interest, shows the same watch pinned at the top
    in a highlighted box marked as the trigger item, followed by recommended
    items that are watches, watch bands and fitness bands.}
  \label{fig:tir}
\end{figure}

In recent years, Generative Recommendation (GR) has emerged as an important paradigm in recommender systems~\cite{rajput2023recommender,10597986,10.1145/3523227.3546767}. Instead of scoring items from a predefined candidate pool, semantic-ID-based GR models directly predict target items by autoregressively generating their discrete Semantic IDs (SIDs)~\cite{rajput2023recommender}. This formulation unifies item representation with generation, showing strong capabilities in modeling complex user behavior sequences. Many representative GR methods formulate recommendation as predicting the next item from user history~\cite{rajput2023recommender,10597986}, a setting we refer to as \textbf{User-Induced Recommendation (UIR)}. In practical search, advertising, and recommendation platforms, another important scenario is \textbf{Trigger-Induced Recommendation (TIR)}~\cite{10.1145/3485447.3511970,10.1145/3543873.3584661,10.1145/3616855.3635829}. As illustrated in Figure~\ref{fig:tir}, when a user interacts with a specific product in a general feed, the platform immediately opens a follow-up feed organized around that item. In this context, the clicked item is no longer an ordinary historical interaction. Instead, it serves as an explicit trigger that conveys the user's immediate shopping intent, which should actively steer the direction of follow-up recommendations.

Although Trigger-Induced Recommendation (TIR) represents a cornerstone scenario across industrial digital platforms, trigger modeling remains largely unexplored in generative recommendation (GR). Existing GR frameworks inherently presume behavioral continuity, lacking dedicated mechanisms to accommodate sudden shifts in user intent. Meanwhile, representative TIR solutions are developed for discriminative CTR prediction using trigger-aware representations~\cite{10.1145/3485447.3511970,10.1145/3543873.3584661,10.1145/3616855.3635829}. Because these methods produce only scalar ranking scores or static representations, they cannot steer discrete, token-by-token autoregressive generation. While extending GR to TIR marks a fundamental transition from discriminative scoring to condition-controlled generation, it inevitably exposes an essential structural tension between historical preference continuity and trigger-driven intent redirection. An overemphasis on continuity reduces the model to a conventional history-based system that fails to capture immediate user impulses. Conversely, an exclusive reliance on the trigger risks confining candidate generation to an overly narrow semantic space, disregarding persistent user preferences and causing severe overspecialization.

Reconciling this structural tension within an autoregressive framework introduces three challenges across encoding, representation, and decoding. First, during sequence encoding, standard self-attention allows extensive interaction histories to easily dilute a single trigger token, preventing it from effectively contextualizing recent user actions without an explicit mechanism to calibrate its influence. Second, the interplay between persistent preferences and immediate intent varies dynamically across requests. A user action may reflect either casual exploration where historical taste should prevail, or a decisive shift toward a new interest requiring strict trigger alignment, making static fusion schemes inadequate. Third, during autoregressive decoding, semantic IDs unfold sequentially from broad categories to fine-grained attributes. Conditioning only at the initial step inevitably weakens over successive generation steps and cannot accommodate the varying semantic granularity of each token, underscoring the need for persistent, position-aware guidance throughout the generation trajectory.

To address these challenges, we propose \textbf{TriGen}, an end-to-end framework that establishes a tailored generative paradigm for trigger-induced recommendation. As illustrated in Figure~\ref{fig:model}, TriGen systematically reconciles preference continuity and intent redirection across the entire autoregressive pipeline through three synergistic designs. At the sequence encoding stage, \textit{Calibrated Tail Injection} places repeated trigger representations at the history tail, using RoPE to model relative positions and the repetition count to regulate trigger influence. At the representation stage, \textit{Adaptive Dual-Path Fusion} decouples persistent historical preferences from immediate trigger impulses, employing an input-dependent gating module to dynamically arbitrate their balance on a per-request basis. Finally, at the autoregressive decoding stage, \textit{Position-Specific Trigger-PFFN} incorporates level-aware transformations directly into the decoder layers, sustaining persistent trigger guidance that adapts to the evolving semantic granularity of each decoding step.

We evaluate TriGen on both a large-scale industrial e-commerce dataset and a public benchmark. Offline experimental results demonstrate that TriGen consistently outperforms competitive sequential, generative, and discriminative TIR baselines. Furthermore, online A/B tests across six international markets achieve traffic-weighted overall lifts of +3.27\% in Revenue and +0.90\% in GMV. TriGen is now fully deployed in production, serving primary online traffic.

The main contributions of this work are summarized as follows:
\begin{itemize}
    \item \textbf{A New Generative Paradigm for TIR:} While TIR has historically relied on discriminative matching methods, our work pioneers an intent-controlled autoregressive sequence generation framework under preference continuity constraints. \textbf{To the best of our knowledge, we are the first to systematically investigate generative recommendation in the TIR scenario.}
    \item \textbf{The TriGen Framework:} We identify the core structural tension between preference continuity and intent redirection in TIR-GR and propose TriGen to resolve it. The framework achieves strength-calibrated local reinterpretation via tail injection with RoPE, adaptively balances dual-path preferences through a gating router, and enforces persistent hierarchical intent guidance during decoding via position-specific TriggerPFFN.
    \item \textbf{Empirical Results and Deployment:} Offline evaluations on large-scale industrial and public datasets demonstrate that TriGen consistently outperforms competitive sequential, generative, and discriminative baselines. Crucially, rigorous online A/B testing across six country markets confirms statistically significant business gains in both Revenue and Gross Merchandise Value, \textbf{culminating in the full deployment of TriGen in production to serve primary online traffic.}
\end{itemize}

\section{Related Work}

\subsection{Trigger-Induced Recommendation}

Industrial recommendation often separates candidate generation from ranking~\cite{10.1145/2959100.2959190}. Within this setting, DIN activates historical behaviors relevant to a target item~\cite{10.1145/3219819.3219823}, while DIEN models the evolution of interests through behavior sequences~\cite{10.1609/aaai.v33i01.33015941}. Sequential recommenders likewise learn preferences from interaction histories using recurrent networks~\cite{hidasi2016sessionbasedrecommendationsrecurrentneural}, causal self-attention~\cite{8594844}, or bidirectional sequence modeling~\cite{10.1145/3357384.3357895}. These approaches provide useful history-modeling mechanisms, but a clicked trigger introduces an additional, explicit signal of immediate intent.

DIHN studies Trigger-Induced Recommendation and highlights historical behaviors through trigger--target fusion and user-intent estimation~\cite{10.1145/3485447.3511970}. DIAN explicitly distinguishes trigger-driven entry from habitual entry and balances trigger-based and trigger-free predictions according to estimated intention~\cite{10.1145/3543873.3584661}. DEI2N further models evolving instant interest using temporal behavior information and trigger--target interactions~\cite{10.1145/3616855.3635829}. These methods motivate the need to coordinate historical preferences with trigger intent. Their original formulations predict CTRs instead of item-token sequences, leaving open how to incorporate trigger information throughout semantic-ID generation.

\subsection{Generative Recommendation}

Transformer models~\cite{Vaswani2017AttentionIA} and unified text-to-text modeling~\cite{3455716.3455856} provide building blocks for sequence generation. P5 formulates multiple recommendation tasks within a unified language-processing framework~\cite{10.1145/3523227.3546767}. In document retrieval, DSI and NCI generate document identifiers~\cite{3600270.3601857,3600270.3602126}, SEAL retrieves documents through generated text substrings~\cite{3600270.3602566}, and GenRet learns discrete document identifiers~\cite{sun2023learningtokenizegenerativeretrieval}. These works establish different ways to connect sequence generation with retrieval; document retrieval and personalized next-item recommendation nevertheless have different conditioning signals and evaluation settings.

For semantic-ID-based recommendation, vector quantization and residual quantization provide discrete representations~\cite{3295222.3295378,9879532}. TIGER represents items with semantic IDs and autoregressively generates the identifier of the next item~\cite{rajput2023recommender}. LC-Rec integrates language and collaborative semantics through item indexing and alignment tuning~\cite{10597986}. LETTER improves item tokenization with semantic, collaborative, and diversity regularization~\cite{10.1145/3627673.3679569}; ETEGRec jointly learns the tokenizer and recommender~\cite{10.1145/3726302.3729989}; and ED$^2$ coordinates dual dynamic indices for sequential recommendation~\cite{10.1145/3696410.3714866}.

Other directions include behavior--semantic collaboration in EAGER~\cite{10.1145/3637528.3671775}, multimodal generation in MMGRec~\cite{10.1145/3838594}, and parallel prediction of long semantic IDs in RPG~\cite{10.1145/3711896.3736979}. RPG demonstrates that semantic-ID recommendation need not use token-by-token autoregressive decoding. At industrial scale, OneRec unifies retrieval and ranking with preference alignment~\cite{deng2025onerecunifyingretrieverank}, while HSTU formulates recommendation as sequential transduction~\cite{3692070.3694484}. Our focus is complementary: we study how an explicit trigger should influence history encoding, preference fusion, and each decoding step in autoregressive semantic-ID recommendation.

\section{Preliminaries}

\subsection{Generative Recommendation with Semantic IDs}
Generative recommendation reframes item recommendation as an autoregressive token generation task. Instead of retrieving items from a large candidate pool, the model directly predicts the discrete tokens of the target item~\cite{rajput2023recommender}.

Formally, each item $i$ is represented by a SID, which is a hierarchical sequence of $L$ discrete tokens:
\begin{equation}
c_i = [c_i^{(1)}, c_i^{(2)}, \dots, c_i^{(L)}], \quad c_i^{(l)} \in \mathcal{V}_l,
\end{equation}
where $c_i^{(l)}$ denotes the token at the $l$-th level, and $\mathcal{V}_l$ is the corresponding codebook vocabulary. Typically obtained via residual quantization methods such as RQ-VAE~\cite{9879532,rajput2023recommender}, these tokens encode successive refinements of an item representation. This motivates level-specific conditioning, without requiring every level to correspond to an explicitly labeled category or attribute.

Let $\mathcal{S}_u = [i_1, i_2, \dots, i_T]$ denote the historical interaction sequence of user $u$. Standard generative recommenders predict the target item $i_{T+1}$ by autoregressively generating its SID tokens conditioned on the history:
\begin{equation}
P(c_{i_{T+1}} \mid \mathcal{S}_u) = \prod_{l=1}^{L} P\left(c_{i_{T+1}}^{(l)} \mid \mathcal{S}_u, c_{i_{T+1}}^{(<l)}\right),
\end{equation}
where $c_{i_{T+1}}^{(<l)} = [c_{i_{T+1}}^{(1)}, \dots, c_{i_{T+1}}^{(l-1)}]$ represents the prefix tokens generated before level $l$.

\subsection{Trigger-Aware Generative Recommendation}
In practical platforms, recommendations are often triggered by a specific user interaction, such as clicking an item in the main feed to enter a related follow-up feed. We define this clicked item as the trigger item $i_{tr}$, represented by its SID $c_{tr} = [c_{tr}^{(1)}, \dots, c_{tr}^{(L)}]$. The trigger reflects the user's immediate interest and should explicitly guide subsequent recommendations.

Unlike standard generative recommendation that relies solely on history, Trigger-Induced Recommendation (TIR) requires the model to generate the target item conditioned on both user history and the trigger item:
\begin{equation}
P(c_{i_{T+1}} \mid \mathcal{S}_u, c_{tr}) = \prod_{l=1}^{L} P\left(c_{i_{T+1}}^{(l)} \mid \mathcal{S}_u, c_{tr}, c_{i_{T+1}}^{(<l)}\right).
\end{equation}
This formulation highlights a core challenge: the trigger $c_{tr}$ must not act as a mere static feature. Instead, it must actively steer the generation trajectory across all decoding levels $l \in \{1, \dots, L\}$ while preserving the user's long-term historical preferences.

\section{Methodology}
\begin{figure*}[t]
  \centering
  \includegraphics[width=\textwidth]{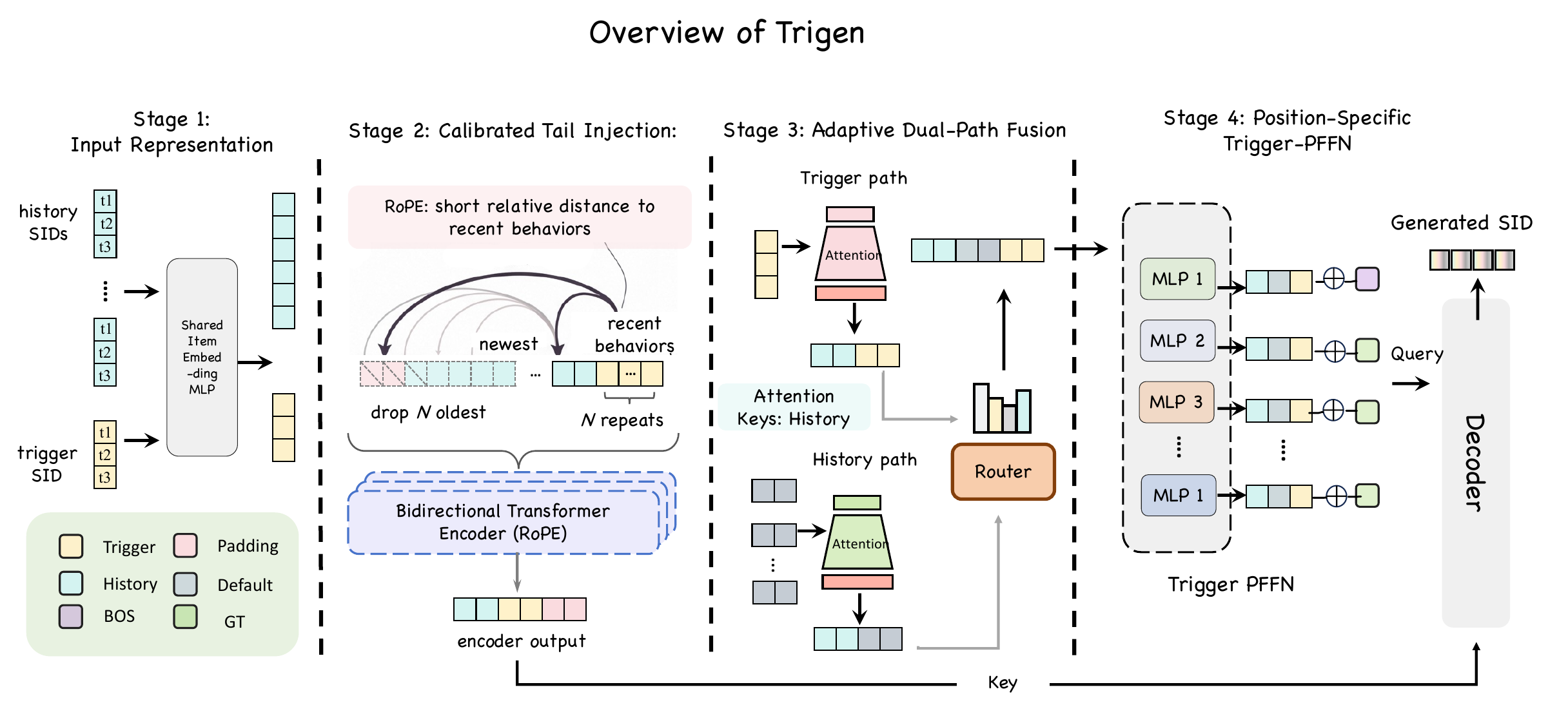}
  \caption{Overall architecture of TriGen. (a) \textbf{Calibrated Tail Injection:} The trigger is appended to the sequence tail with $N$ repetitions, placing it adjacent to recent behaviors. RoPE models relative positions, while $N$ controls the repetition count to regulate
trigger influence. (b)~\textbf{Adaptive Dual-Path Fusion}: Parallel pathways decouple trigger-conditioned steering from history-preserved preference, dynamically balanced by an input-dependent gating router. (c)~\textbf{Position-Specific Trigger-PFFN}: Step-specific feed-forward networks project the fused representation into level-aware condition signals, maintaining persistent, coarse-to-fine guidance throughout autoregressive decoding.}
  \Description{Schematic diagram of the TriGen framework, comprising three core stages from sequence encoding to autoregressive decoding. In the sequence encoding stage, Semantic IDs of historical behaviors and the trigger are mapped into a shared latent space; the trigger tokens are appended to the sequence tail with calibrated repetition and encoded via a Transformer layer with Rotary Position Embedding. In the representation stage, the representations bifurcate into a trigger-guided branch and a history-preserving branch, whose outputs are dynamically combined via a gating router. In the decoding stage, the fused representation is transformed by position-specific feed-forward layers to supply level-aware conditional signals to each step of the autoregressive decoder as it generates target Semantic ID tokens.}
  \label{fig:model}
\end{figure*}

To systematically reconcile preference continuity and intent redirection, we propose TriGen, an encoder-decoder generative framework tailored for trigger-induced recommendation. As illustrated in Figure~\ref{fig:model}, TriGen operationalizes this trade-off across the entire generation pipeline through three synergistic designs. At the sequence encoding stage, \textbf{Calibrated Tail Injection} places repeated trigger representations at the history tail, using RoPE to model relative positions and the repetition count to regulate trigger influence. At the representation stage, \textbf{Adaptive Dual-Path Fusion} decouples persistent historical preferences from immediate trigger impulses, employing an input-dependent gating module to dynamically balance both paths for each request. Finally, at the autoregressive decoding stage, \textbf{Position-Specific Trigger-PFFN} incorporates level-aware transformations directly into the decoder layers, sustaining persistent trigger guidance that adapts to the evolving semantic granularity of each decoding step. We elaborate on each component below.



\subsection{Calibrated Tail Injection}

To address local reinterpretation and strength control in TIR, we propose \textit{Calibrated Tail Injection}, which incorporates the trigger directly into the sequence tail within encoder self-attention. The core intuition is that an immediate trigger click does not reshape the entire interaction history uniformly; rather, it primarily modulates a localized subset of behaviors, particularly recent and relevant ones. Placing the trigger inside the sequence rather than treating it as an external global condition enables direct attention interactions with this relevant local context.

Formally, let $H = (h_1, h_2, \dots, h_M)$ denote the historical sequence representation within a fixed context window of length $M$, where each $h_j \in \mathbb{R}^d$, and let $e^{trg} \in \mathbb{R}^d$ be the representation of the trigger item. To explicitly calibrate the trigger's intervention strength over historical behaviors, we introduce $N$ consecutive copies of the trigger at the tail of the sequence, yielding:
\begin{equation}
\tilde{H} = (h_{N+1}, \dots, h_M, \underbrace{e^{trg}, \dots, e^{trg}}_{N \text{ times}}).
\end{equation}

This design coordinates two complementary mechanisms. Placing the trigger at the sequence tail makes it positionally adjacent to the user's most recent behaviors. Combined with the relative-position modeling of Rotary Position Embedding (RoPE)~\cite{10.1016/j.neucom.2023.127063}, this arrangement is intended to facilitate interactions between the trigger and recent behaviors. Meanwhile, the repetition factor $N$ regulates the query-key capacity occupied by the trigger in self-attention: a larger $N$ empowers the trigger to steer the context toward immediate intent, whereas a smaller $N$ prevents it from overriding historical preferences.

Finally, $\tilde{H}$ is processed by a bidirectional Transformer encoder to produce the trigger-aware contextual sequence $O = (o_1, o_2, \dots, o_M)$, which provides representations for both the Adaptive Dual-Path Fusion module and the decoder cross-attention.

\subsection{Adaptive Dual-Path Fusion}
\label{sec:dual_path_fusion}

While Calibrated Tail Injection regulates the trigger's localized intervention strength within encoder self-attention, the resulting representations inevitably couple persistent historical preferences with immediate trigger impulses. Disentangling these two factors is crucial in trigger-induced recommendation, as user behavior may reflect both habitual preferences and trigger-induced intent~\cite{10.1145/3543873.3584661}. If the system over-emphasizes the trigger item, recommendations risk collapsing into myopic over-reaction or being misled by accidental clicks; conversely, if it adheres strictly to historical habits, the redirective signal of the trigger is effectively lost. Consequently, the generative condition must be explicitly decoupled and adaptively reconciled on a per-request basis. To this end, we propose Adaptive Dual-Path Fusion (Figure~\ref{fig:model}, Stage~3), which decouples condition modeling into a trigger-guided instant path and a history-preserving baseline path, dynamically arbitrating their synthesis via an input-dependent gating router.

\paragraph{Decoupled Dual-Path Condition Modeling}
To prevent mutual interference between transient impulses and baseline habits, we bypass the coupled encoder sequence and extract two complementary condition pathways directly from the unperturbed historical representations. Specifically, the trigger path projects the trigger item into distinct step-specific queries to retrieve historical behaviors relevant to the immediate intent. Symmetrically, the history path uses a set of learnable default queries (marked as ``Default'' in Figure~\ref{fig:model}) to summarize the user's intrinsic preference prototypes independently of the trigger. Formally, for each decoding step, the condition vectors from both paths are extracted through attention over the raw historical sequence:
\begin{equation}
\label{eq:dual_path_extraction}
\begin{aligned}
t_l &= \text{Softmax}\left(\frac{(\mathbf{W}_l^{tr} e^{trg}) (H \mathbf{W}_K)^\top}{\sqrt{d}}\right) (H \mathbf{W}_V), \\
h_l &= \text{Softmax}\left(\frac{q_l^{def} (H \mathbf{W}_K)^\top}{\sqrt{d}}\right) (H \mathbf{W}_V),
\end{aligned}
\end{equation}
where $l \in \{1, \dots, L\}$ indexes the decoding step, $e^{trg} \in \mathbb{R}^d$ is the trigger representation, and $H \in \mathbb{R}^{M \times d}$ denotes the pristine historical sequence. Here, $\mathbf{W}_l^{tr} \in \mathbb{R}^{d \times d}$ is the step-specific projection matrix, $q_l^{def} \in \mathbb{R}^d$ is the learnable default query, and $\mathbf{W}_K, \mathbf{W}_V \in \mathbb{R}^{d \times d}$ project historical behaviors into shared key and value spaces. Stacking these vectors yields the trigger condition sequence $T = [t_1, \dots, t_L]$ and the history condition sequence $H_{cond} = [h_1, \dots, h_L]$, capturing immediate redirection and baseline continuity.

\paragraph{Hierarchical Gated Routing}
Because the intensity of intent redirection naturally varies across requests, a static combination of both paths is inadequate. We introduce an input-dependent router to learn dynamic fusion weights on a per-request basis. The router first condenses each pathway into a global profile via sequence-level mean pooling, computes step-wise gating coefficients, and synthesizes the calibrated condition vector through a position-wise convex combination:
\begin{equation}
\label{eq:gated_routing}
\begin{aligned}
\mathbf{w} &= [w_1, \dots, w_L] = \sigma\left(\mathbf{W}_r \left[\frac{1}{L}\sum_{j=1}^L t_j \;\|\; \frac{1}{L}\sum_{j=1}^L h_j\right] + \mathbf{b}_r\right), \\
c_l &= w_l t_l + (1 - w_l) h_l, \quad \forall l \in \{1, \dots, L\},
\end{aligned}
\end{equation}
where $[\cdot \,\|\, \cdot]$ denotes vector concatenation, $\mathbf{W}_r \in \mathbb{R}^{L \times 2d}$ and $\mathbf{b}_r \in \mathbb{R}^L$ are learnable routing parameters, and $\sigma(\cdot)$ is the element-wise sigmoid function. The resulting sequence $C = [c_1, \dots, c_L] \in \mathbb{R}^{L \times d}$ establishes a unified condition signal that balances instant redirection with historical continuity, which is subsequently passed to Stage~4 to steer autoregressive generation.

\subsection{Position-Specific Trigger-PFFN}
\label{sec:trigger_pffn}

While encoder-side representations and dual-path condition fusion offer essential global constraints, steering the target generation also requires step-level guidance on the decoder side. In generative recommendation, the target item is represented as a Semantic ID and generated token by token in an autoregressive manner. Residual-quantized Semantic IDs encode an item representation through successive quantization levels~\cite{9879532,rajput2023recommender}. Earlier tokens provide an initial approximation, while later tokens encode residual refinements; their semantic roles can therefore differ across positions. Consequently, different decoding steps naturally require distinct forms of trigger modulation instead of a uniform signal across all positions. To provide step-tailored guidance throughout the generative process, we introduce the Position-Specific Trigger Feed-Forward Network (Trigger-PFFN) on the decoder side (Figure~\ref{fig:model}, Stage~4).

Taking the calibrated condition sequence $C = [c_1, \dots, c_L] \in \mathbb{R}^{L \times d}$ from Stage~3 as input, Trigger-PFFN first maps each condition vector $c_l$ through a dedicated feed-forward layer to adapt to the semantic granularity of the corresponding position:
\begin{equation}
\label{eq:pffn_projection}
\tilde{c}_l = \mathrm{MLP}_l(c_l), \quad l \in \{1, \dots, L\},
\end{equation}
where $\mathrm{MLP}_l(\cdot)$ is an independent multi-layer perceptron learned specifically for the $l$-th Semantic ID level. By using independent projection parameters for each position, the model can flexibly tailor the trigger condition, providing broad categorical guidance for early decoding steps and detailed attribute control for later steps.

To maintain continuous guidance across the entire generation process, we incorporate the modulated condition vectors $\tilde{C} = [\tilde{c}_1, \dots, \tilde{c}_L] \in \mathbb{R}^{L \times d}$ directly into the decoder input representations. Under the standard teacher-forcing protocol with the ground-truth Semantic ID sequence $(y_1, y_2, \dots, y_L)$, the right-shifted decoder input at step $l$ is formed as:
\begin{equation}
\label{eq:decoder_input}
x_l^{dec} = 
\begin{cases}
e(BOS) + \tilde{c}_1, & l = 1, \\
e(y_{l-1}) + \tilde{c}_l, & l \in \{2, \dots, L\},
\end{cases}
\end{equation}
where $e(\cdot) \in \mathbb{R}^d$ denotes the token embedding lookup function and $BOS$ represents the start-of-sequence token. By adding the position-specific condition vector $\tilde{c}_l$ to the input at each position, the trigger guidance accompanies the target prefix at every decoding step. The resulting sequence $X^{dec} = [x_1^{dec}, \dots, x_L^{dec}] \in \mathbb{R}^{L \times d}$ is then processed by causal self-attention and cross-attention over the encoder output $O \in \mathbb{R}^{M \times d}$, enabling the decoder to autoregressively predict the target tokens with hierarchy-aware trigger steering.

\section{Experiments}

\begin{table*}[t]
  \centering
  \caption{Overall performance comparison on the industrial dataset and the Amazon \textit{Industrial\_and\_Scientific} benchmark. Results report the mean performance over 5 seeds (corresponding standard deviations are detailed in Appendix~\ref{sec:appendix_stds}, Table~\ref{tab:stds_results}). R@$K$ and N@$K$ denote Recall@$K$ and NDCG@$K$, respectively. The best performance in each column is highlighted in \textbf{bold}, and the second-best is \underline{underlined}.}
  \label{tab:main_results}
  \begin{tabularx}{\textwidth}{@{}ll*{8}{>{\centering\arraybackslash}X}@{}}
    \toprule
    \multirow{2}{*}{Category} & \multirow{2}{*}{Method} &
    \multicolumn{4}{c}{Industrial Dataset} &
    \multicolumn{4}{c}{Amazon \textit{Industrial\_and\_Scientific}} \\
    \cmidrule(lr){3-6} \cmidrule(lr){7-10}
    & & R@5 & R@10 & N@5 & N@10 & R@5 & R@10 & N@5 & N@10 \\
    \midrule
    \multirow{3}{*}{TIR-specific}
      & DIHN                          & 1.54 & 6.39 & 1.08 & 2.63 & 4.78 & 6.83 & 3.03 & 3.69 \\
      & DIAN                          & 1.59 & 6.80 & 1.15 & 2.82 & 4.93 & 6.94 & 3.64 & 4.26 \\
      & DEI2N                         & 1.61 & 7.20 & 1.19 & 2.98 & 5.17 & 7.46 & 4.08 & 4.81 \\
    \midrule
    \multirow{2}{*}{Sequential}
      & GRU4Rec                       & 1.66 & 7.26 & 1.22 & 3.02 & 7.32 & 9.97 & 6.03 & 6.88 \\
      & SASRec                        & 2.16 & 7.84 & 1.57 & 3.39 & 9.34 & 11.88 & 7.19 & 8.01 \\
    \midrule
    \multirow{3}{*}{\shortstack[l]{Generative\\(Trigger-free)}}
      & HSTU                          & 2.58 & 7.13 & 1.83 & 3.28 & 9.35 & 11.64 & 7.18 & 7.91 \\
      & TIGER                         & 2.39 & 7.53 & 1.74 & 3.38 & 9.75 & 12.41 & 7.51 & 8.36 \\
      & LC-Rec                        & 2.46 & 7.30 & 1.79 & 3.34 & 10.20 & 12.98 & 7.95 & 8.85 \\
    \midrule
    \multirow{3}{*}{Simple TIR-GR}
      & GR + trigger append           & 4.35 & \underline{12.15} & 3.12 & \underline{5.62} & \underline{10.27} & \textbf{13.26} & 7.93 & 8.89 \\
      & GR + trigger conditioning     & \underline{4.46} & 11.44 & \underline{3.28} & 5.51 & 10.24 & 12.99 & 7.85 & 8.74 \\
      & GR + trigger-history attention& 3.33 & 9.80 & 2.41 & 4.47 & 6.81 & 9.38 & 4.72 & 5.54 \\
    \midrule
    Ours
      & \textbf{TriGen}               & \textbf{5.58} & \textbf{15.50} & \textbf{4.16} & \textbf{7.33} & \textbf{10.83} & \underline{13.22} & \textbf{8.45} & \textbf{9.21} \\
    \bottomrule
  \end{tabularx}
\end{table*}

\subsection{Experimental Setup}

\subsubsection{Datasets}

We use industrial and public datasets. The industrial dataset is collected from a real Trigger-Induced Recommendation scenario on a major e-commerce platform. It contains large-scale user behavior and feedback logs. The dataset covers real exposure, click, and follow-up feedback signals, supporting industrial TIR evaluation.

For the public benchmark, we use the \textit{Industrial\_and\_Scientific} category from Amazon Reviews 2018~\cite{ni-etal-2019-justifying}. We use the 5-core version and keep interactions from 2016-10 to 2018-11. Because this public dataset does not provide native TIR annotations, we construct simulated triggers on it. Specifically, we treat the most recent interaction in the user history as the trigger and use it to predict the subsequent target item. This assumes that the most recent interaction reflects instant interest. The statistics of the public dataset are reported in Table~\ref{tab:data}.

\subsubsection{Evaluation Metrics}

We use Recall@K and NDCG@K~\cite{10.1145/582415.582418} as offline evaluation metrics, where $K \in \{5,10\}$. The model first generates the Semantic ID sequence of the target item and then maps it back to the final item. For online evaluation, we report business results in six country markets, namely Indonesia, Malaysia, Singapore, the Philippines, Thailand, and Vietnam. In addition, we report the weighted aggregated GMV and REV across these six countries.

\subsubsection{Compared Methods}
\label{sec:compared_methods}

To evaluate the effectiveness of TriGen and investigate whether naive trigger conditioning suffices, we compare our framework against four categories of baselines:
(1) \textbf{TIR-specific discriminative methods}, including DIHN~\cite{10.1145/3485447.3511970}, DEI2N~\cite{10.1145/3616855.3635829}, and DIAN~\cite{10.1145/3543873.3584661}, which model trigger information under matching or CTR-scoring paradigms;
(2) \textbf{Standard sequential recommendation methods}, including GRU4Rec~\cite{hidasi2016sessionbasedrecommendationsrecurrentneural} and SASRec~\cite{8594844}, which rely solely on historical interaction sequences without trigger modeling;
(3) \textbf{Generative and sequential-transduction baselines}, including TIGER~\cite{rajput2023recommender}, HSTU~\cite{3692070.3694484}, and LC-Rec~\cite{10597986}; and
(4) \textbf{Simple TIR-GR baselines}, comprising \textit{GR + trigger append}, \textit{GR + trigger conditioning}, and \textit{GR + trigger-history attention}. These baselines adopt the identical generative backbone, codebook, and parameter budget as TriGen to examine whether straightforward trigger injection can deliver comparable gains.

Detailed configurations of all compared methods are provided in Appendix~\ref{appendix:baselines}.







\subsubsection{Implementation Details}

All methods are trained and evaluated on the same data split. We use AdamW~\cite{loshchilov2018decoupled} for optimization and beam search for inference. Semantic IDs are fixed discrete codes and are not updated during training. Configurations are provided in Appendix~\ref{app:impl}.

\subsection{Overall Performance}

Table~\ref{tab:main_results} compares all methods on the industrial and Amazon datasets. Overall, TriGen achieves the best or highly competitive performance on both datasets. These results support explicit trigger modeling in generative recommendation.

Compared with TIR-specific methods, TriGen directly models how the trigger affects hierarchical Semantic ID generation within a generative framework, combining instant interest with generation. Compared with standard sequential recommendation methods such as GRU4Rec and SASRec, TriGen's gains indicate that history alone is insufficient to capture current intent in TIR. Compared with generative recommendation baselines such as TIGER, HSTU, and LC-Rec, the consistent gains of TriGen suggest that TIR also requires dedicated trigger modeling.

Crucially, TriGen consistently outperforms the Simple TIR-GR baselines across both datasets. Naive strategies such as GR + trigger append and GR + trigger conditioning fail to calibrate trigger influence against long histories and cannot accommodate the coarse-to-fine granularity of Semantic IDs, limiting gains under noisy triggers. The decisive margin between TriGen and these simple baselines rules out the hypothesis that any arbitrary trigger conditioning scheme suffices, confirming that our tailored architectural designs provide benefits far beyond the mere presence of trigger information.

\subsection{Ablation Study}
\label{sec:ablation_study}

To examine the contributions of TriGen's components, we evaluate the full model against four variants on the industrial dataset under identical backbone configurations:
(1) \textbf{no trigger}, which removes the trigger item entirely and relies solely on user history for recommendation;
(2) \textbf{w/o Calibrated Tail Injection}, which excludes the trigger from the encoder historical sequence;
(3) \textbf{w/o Adaptive Dual-Path Fusion}, which removes the dual-path condition channels and the gating router, and instead feeds the original trigger embedding into the retained position-specific Trigger-PFFN; and
(4) \textbf{w/o Position-Specific Trigger-PFFN}, which removes the position-specific trigger projections and completely eliminates trigger conditioning from the decoder.

\begin{table}[t]
  \centering
  \caption{Ablation study of TriGen on the industrial dataset. R@$K$ and N@$K$ denote Recall@$K$ and NDCG@$K$ (\%), respectively. The best performance in each column is highlighted in \textbf{bold}.}
  \label{tab:ablation}
  \small
  \begin{tabular}{lcccc}
    \toprule
    Variant & R@5 & R@10 & N@5 & N@10 \\
    \midrule
    no trigger                            & 2.88 & 8.27 & 2.08 & 3.81 \\
    w/o Calibrated Tail Injection         & 4.16 & 11.04 & 2.98 & 5.19 \\
    w/o Adaptive Dual-Path Fusion         & 5.11 & 14.51 & 3.61 & 6.34 \\
    w/o Position-Specific Trigger-PFFN    & 4.28 & 11.10 & 3.11 & 5.31 \\
    \textbf{TriGen} (full)                & \textbf{5.58} & \textbf{15.50} & \textbf{4.16} & \textbf{7.33} \\
    \bottomrule
  \end{tabular}
\end{table}

As shown in Table~\ref{tab:ablation}, the full TriGen model performs best across all metrics, supporting the contribution of each component.

Removing the trigger causes the largest accuracy drop. In trigger-induced scenarios, user behavior is strongly governed by the immediate interaction. Without the trigger, the model cannot identify the user's instant intent, demonstrating that historical interaction data alone cannot yield accurate recommendations in such contexts.

Excluding the trigger from sequence encoding also degrades recommendation accuracy substantially. This indicates that relying solely on downstream trigger conditioning is insufficient. By appending the trigger to the sequence tail, the encoder exploits self-attention to contextualize past behaviors against the trigger, particularly the most recent interactions. Omitting this injection leaves historical representations uninformed by the trigger context.

Removing Adaptive Dual-Path Fusion while retaining
Trigger-PFFN reduces both Recall and NDCG. In this variant, the decoder-side condition is derived directly from the original trigger embedding instead of from the fused trigger-guided and history-preserving representations. The performance gap supports the benefit of incorporating historical context through adaptive dual-path fusion, compared with using the trigger embedding alone.

Finally, removing Trigger-PFFN and eliminating all trigger conditioning from the decoding stage degrades recommendation performance. Compared with this variant, the full model achieves better results by providing position-specific trigger guidance throughout autoregressive generation. These results support the effectiveness of explicit
decoder-side trigger conditioning.

\subsection{Fine-grained Design Analysis}
\label{sec:design_analysis}

\begin{figure}[t]
  \centering
  \includegraphics[width=\columnwidth]{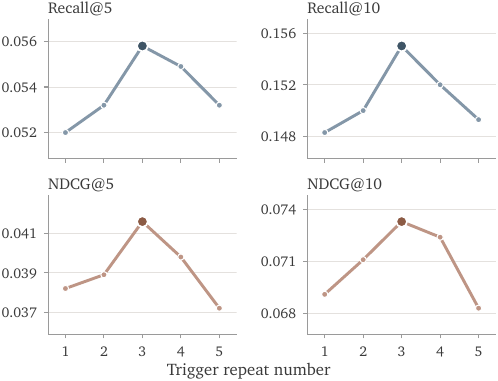}
  \caption{Effect of the trigger repeat number on the industrial dataset. All evaluation metrics first improve and then decline, confirming that a calibrated repetition count achieves the best balance.}
  \Description{Four line charts arranged in a two-by-two grid, one each for
    Recall at 5, Recall at 10, NDCG at 5 and NDCG at 10 on the industrial
    dataset, with the trigger repeat number from one to five on the shared
    horizontal axis. In every chart the metric first increases and then
    decreases, peaking at a repeat number of three rather than at the
    smallest or the largest value.}
  \label{fig:repeat}
\end{figure}

We further examine architectural choices within each component. Industrial-dataset results are reported in Table~\ref{tab:design} and Figure~\ref{fig:repeat}.

\textbf{Encoder-side Designs.} We first examine the role of Calibrated Tail Injection. Placing the trigger at the sequence tail outperforms placing it at the sequence head. Tail placement makes the trigger positionally adjacent to recent behaviors, and the observed improvement supports this placement choice in our setting.

Removing rotary position embeddings leads to a performance decline, supporting the effectiveness of RoPE in the trigger-injected encoder. Together with tail placement, RoPE provides relative-position information for modeling interactions between the trigger and historical behaviors.

Finally, excluding repeated insertion reduces accuracy. Inserting the trigger only once fails to provide sufficient intervention strength within the sequence, while Figure~\ref{fig:repeat} shows that performance first increases and then gradually declines as the repetition count grows. This bell-shaped trend demonstrates that trigger influence must be properly calibrated: an insufficient count leaves the trigger overshadowed by long history, whereas excessive repetition overwhelms the user's historical context.

\textbf{Dual-Path Fusion Designs.} We further investigate the design of the gating module in Adaptive Dual-Path Fusion by comparing it with a static uniform combination, where both paths are simply averaged without routing weights.

As shown in Table~\ref{tab:design}, replacing the adaptive router with uniform averaging results in a clear drop in performance. This degradation demonstrates that different requests require different balances between instant redirection and historical continuity. An incidental trigger click should rely more heavily on baseline historical tastes, while an exploratory session demands stronger trigger guidance. By dynamically predicting the routing weights from the global states of both pathways, the gating module effectively tailors the condition to individual user requests.

\textbf{Decoder-side Designs.} We evaluate two key design aspects of Position-Specific Trigger-PFFN on the decoder side: the injection position and the projection parameterization.

Comparing full position-wise injection against injecting the condition exclusively at the start token reveals that start-only injection is inadequate for autoregressive generation. In a multi-step decoding process, introducing the condition only at the initial position weakens its influence as the generated prefix grows longer. Providing condition guidance at every decoding step ensures persistent steering throughout the entire target item generation.

Moreover, replacing position-specific projections with a single shared feed-forward layer noticeably hurts generation quality. Because Semantic IDs represent items from broad categories down to specific attributes, each decoding position is responsible for a distinct semantic granularity. Dedicated projections allow early positions to focus on category-level steering while letting later positions specialize in attribute-level details, matching the coarse-to-fine structure of Semantic IDs.

\begin{table}[t]
  \centering
  \caption{Fine-grained design analysis on the industrial dataset. All variants share the same backbone architecture and differ only in the evaluated design choice. R@$K$ and N@$K$ denote Recall@$K$ and NDCG@$K$ (\%), respectively. The best performance in each column is highlighted in \textbf{bold}.}
  \label{tab:design}
  \small
  \begin{tabular}{lcccc}
    \toprule
    Setting & R@5 & R@10 & N@5 & N@10 \\
    \midrule
    \rowcolor[rgb]{0.88,0.93,0.98}
    \multicolumn{5}{l}{\textbf{Encoder-side Designs}} \\
    trigger at sequence head          & 5.31 & 14.58 & 3.88 & 6.79 \\
    w/o RoPE                          & 4.96 & 14.24 & 3.58 & 6.56 \\
    w/o repeat                        & 5.20 & 14.83 & 3.82 & 6.91 \\
    \addlinespace
    \rowcolor[rgb]{0.88,0.93,0.98}
    \multicolumn{5}{l}{\textbf{Dual-Path Fusion Designs}} \\
    uniform path combination          & 4.15 & 11.87 & 2.98 & 5.47 \\
    \addlinespace
    \rowcolor[rgb]{0.88,0.93,0.98}
    \multicolumn{5}{l}{\textbf{Decoder-side Designs}} \\
    start-only injection              & 4.62 & 12.93 & 3.36 & 5.99 \\
    shared projection layer           & 4.90 & 13.85 & 3.61 & 6.51 \\
    \addlinespace
    \textbf{TriGen} (full)            & \textbf{5.58} & \textbf{15.50} & \textbf{4.16} & \textbf{7.33} \\
    \bottomrule
  \end{tabular}
\end{table}

\subsection{Online A/B Test}
\label{sec:online_ab_test}

\begin{figure}[t]
  \centering
  \includegraphics[width=\columnwidth]{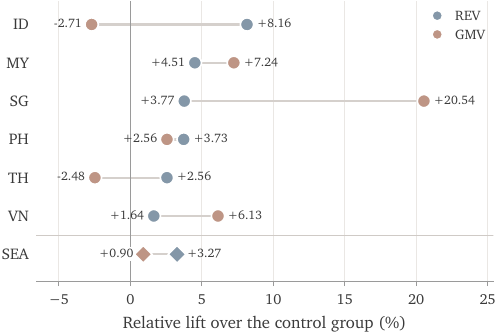}
  \caption{Online A/B test results across six international markets. REV (Revenue) and GMV (Gross Merchandise Value) denote the relative lifts of TriGen over the deployed production baseline. SEA represents the traffic-weighted aggregate.}
  \Description{Dot plot of the online A/B test lifts, showing relative improvements across six country markets and the traffic-weighted aggregate SEA.}
  \label{fig:ab}
\end{figure}

To evaluate TriGen in production, we conducted an online A/B test on an e-commerce platform across six country markets: Indonesia (ID), Malaysia (MY), Singapore (SG), the Philippines (PH), Thailand (TH), and Vietnam (VN). The experiment was deployed within the trigger-induced feed on product detail pages, a surface with exposure, click, and conversion signals. We adopted the deployed production model, which implements the competitive generative baseline \textit{GR + trigger append}, as the control group, and deployed TriGen as the treatment group. The evaluation ran from June 28 to July 2, 2026, with each bucket allocated 10\% of randomly sampled production traffic, while upstream candidate sources and downstream serving pipelines remained strictly identical.

The results are summarized in Figure~\ref{fig:ab}. TriGen demonstrates robust performance improvements across heterogeneous international markets. In terms of platform monetization, TriGen achieves consistent revenue gains across all six countries, yielding relative lifts between +1.64\% and +8.16\%. Gross Merchandise Value (GMV) concurrently increases in four out of six countries, achieving up to a +20.54\% improvement. On the traffic-weighted aggregate across Southeast Asia (SEA), TriGen delivers an overall lift of +3.27\% in Revenue and +0.90\% in GMV, with metrics demonstrating strong statistical significance ($p < 0.05$, two-sided test).

During serving, the additional computational overhead of TriGen is practically negligible. Because the tailored components rely solely on lightweight cross-attentions, a compact routing unit, and compact position-specific feed-forward layers, the online inference latency is approximately 10.5\,ms, remaining virtually indistinguishable from that of the control model. These production results substantiate our offline findings, confirming that resolving the structural tension between trigger steering and preference continuity brings decisive commercial value under strict serving budgets. Following the sustained performance advantages observed during the trial, TriGen has been fully deployed to production since July 2026, serving primary online traffic across the platform.

\subsection{Case Analysis}

\begin{figure}
  \centering
  \includegraphics[width=\columnwidth]{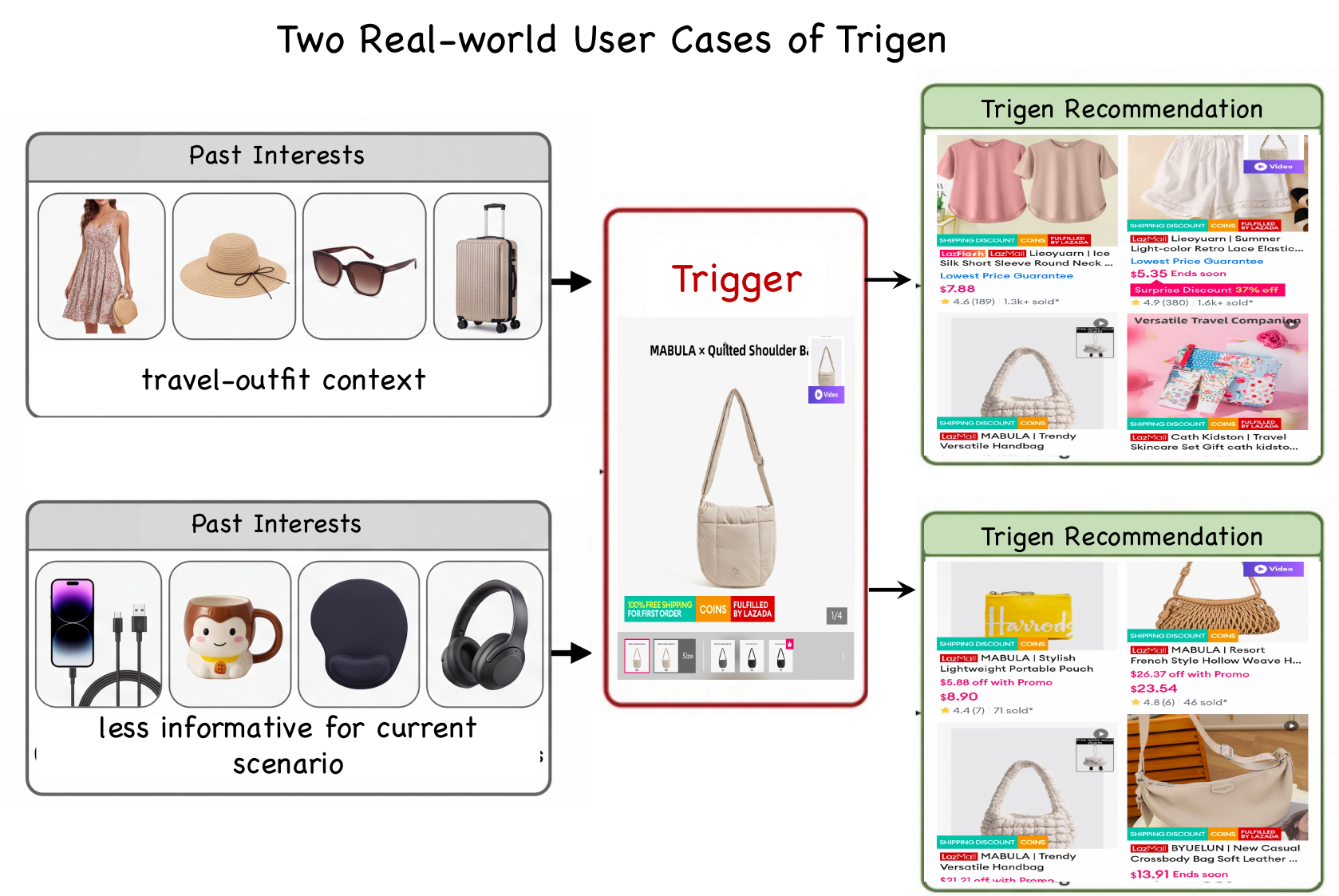}
  \caption{Two real-world cases that share the same trigger, a quilted
    shoulder bag. In the top case the history provides a travel-outfit
    context; in the bottom case it is unrelated to the trigger. Each row
    shows the history, the trigger, and the items TriGen generates.}
  \Description{Two recommendation cases arranged as two rows, both pointing
    to the same trigger item shown in the middle, a quilted shoulder bag. The
    top row starts from a history of travel outfit items, a dress, a sun hat,
    sunglasses and a suitcase, and TriGen returns summer tops, lace shorts, a
    handbag and a travel skincare set. The bottom row starts from a history
    unrelated to the trigger, a phone with a charging cable, a mug, a mouse
    pad and headphones, and TriGen returns four bags: a pouch, a woven resort
    bag, a handbag and a crossbody bag.}
  \label{fig:case}
\end{figure}

The offline and online results show that TriGen improves accuracy and business metrics, but they do not directly show how the trigger affects generation. We therefore inspect two real-world users in Figure~\ref{fig:case} who clicked the same trigger, a quilted shoulder bag, but have different histories.

In the first case, the history is informative. The user has browsed a dress, a sun hat, sunglasses, and a suitcase, which form a travel-outfit context. TriGen does not generate only bags. It returns summer tops and lace shorts that continue the outfit line, one bag close to the trigger, and a travel skincare set. This suggests that the history-preserving path keeps the travel scenario available, so the trigger is interpreted as part of the trip.

In the second case, the history provides little support for the click. The recent items are digital and desk accessories, which are unrelated to a shoulder bag. Here TriGen concentrates all slots on bags and varies style and material instead of category. Without useful history context, the model relies more on the trigger-conditioned path. Together, the two cases show that the fusion is adaptive. The same trigger is expanded into a broader scenario when the history supports it, and followed more closely when it does not.

\section{Conclusion}

In this work, we study generative recommendation for Trigger-Induced Recommendation. To model the user's instant interest carried by the trigger, we propose TriGen, which combines encoder-side trigger injection, DualPathFusion, and TriggerPFFN. Experiments on both industrial and public datasets show that TriGen consistently outperforms strong baselines. Online A/B testing further validates its effectiveness in a real production environment, and TriGen has now been fully deployed online. Overall, our results show that explicit trigger modeling is crucial for generative recommendation in TIR.

\bibliographystyle{ACM-Reference-Format}
\bibliography{trigen_references_true}

\appendix

\section{Dataset and Trigger Construction}
\label{app:data}

Trigger-induced recommendation is usually evaluated on the Alimama dataset, but its fields are anonymized so aggressively that no usable item text survives, and without text there is nothing for the Semantic ID tokenizer of Appendix~\ref{app:sid} to quantize. We therefore evaluate on Amazon Reviews~\cite{ni-etal-2019-justifying}, a widely used public benchmark in generative recommendation~\cite{rajput2023recommender,10597986}, and simulate the trigger as described below.

The public dataset is the \textit{Industrial\_and\_Scientific} category of Amazon Reviews 2018, in its 5-core version and restricted to the window from 2016-10 to 2018-11. Every user's chronologically ordered interactions are expanded into (history, target) samples by a sliding window whose history keeps the $10$ most recent items, and the samples of each user are split chronologically at a ratio of about $8{:}1{:}1$, so the earliest windows form the training set and the latest ones the test set. Item text is used only by the tokenizer of Appendix~\ref{app:sid}; the recommenders consume Semantic IDs only. Statistics of the industrial dataset are reported in Table~\ref{tab:ind_data}.

The public data carries no trigger field, so the trigger has to be simulated. Given a sample with history $(x_1,\dots,x_m)$ and target $y$, we take the most recent history item as the trigger and keep the rest as the long-term history:
\begin{equation}
  (x_1,\dots,x_{m-1},x_m),\ y \ \Longrightarrow\ \underbrace{(x_1,\dots,x_{m-1})}_{\text{history}},\ \underbrace{x_m}_{\text{trigger}},\ y ,
\end{equation}
which rests on the assumption that the most recent interaction explicitly guides the next one. This split only changes how the sequence is presented, not what it contains. Models that consume a trigger read $(x_1,\dots,x_{m-1})$ as history and $x_m$ as the trigger, while every trigger-free baseline reads the whole sequence $(x_1,\dots,x_m)$ as its history. Both therefore see exactly the same items, and no baseline is shortened by the rewriting. Samples whose original history has length one are dropped for all models alike, since they would leave the history-preserving path of DualPathFusion without input, so the evaluated sample set stays identical across models. Table~\ref{tab:data} reports the statistics both before and after this step. Aggregated over the three splits, the data contain 3,686 items, 7,694 users and 49,371 interactions, i.e.\ a density of 0.174\%.

\begin{table}[!t]
  \caption{Statistics of the industrial dataset.}
  \label{tab:ind_data}
  \begin{tabular}{@{}lrrr@{}}
    \toprule
    Dataset & \#Users & \#Items & \#Samples \\
    \midrule
    Industrial & 49{,}277{,}990 & 158{,}533{,}241 & 1{,}492{,}770{,}272 \\
    \bottomrule
  \end{tabular}
\end{table}

\begin{table}[!t]
  \caption{Statistics of the public dataset before and after the trigger rewriting. ``After'' is the sample set used by every model; avg. $|H|$ there excludes the trigger, which trigger-free baselines still read as the last history item.}
  \label{tab:data}
  \begin{tabular}{@{}llrrr@{}}
    \toprule
    & Statistic & Train & Valid & Test \\
    \midrule
    \multirow{4}{*}{\rotatebox{90}{before}}
      & samples      & 36{,}259 & 4{,}532 & 4{,}533 \\
      & users        &  7{,}034 & 2{,}011 & 1{,}987 \\
      & avg. $|H|$   &     3.77 &    4.64 &    4.96 \\
      & max. $|H|$   &       10 &      10 &      10 \\
    \midrule
    \multirow{4}{*}{\rotatebox{90}{after}}
      & samples      & 29{,}225 & 4{,}139 & 4{,}266 \\
      & dropped      &  7{,}034 &    393  &    267  \\
      & avg. $|H|$   &     3.44 &    3.99 &    4.20 \\
      & max. $|H|$   &        9 &       9 &       9 \\
    \bottomrule
  \end{tabular}
\end{table}

One consequence of this protocol deserves an explicit statement, because it works against TriGen rather than for it. In production the trigger is a real event: the user has just clicked the item, and that click carries a short-term intent that the next interaction often follows. The simulated trigger has no such event behind it. It is an ordinary past interaction that merely happens to be the most recent one, so the intent signal TriGen is designed to exploit is weaker and noisier here than in deployment. Because the baselines keep the full sequence, the public-dataset margin is best read as a lower bound on what trigger modelling contributes: it is obtained with a degraded trigger, and the industrial dataset, where the trigger is a genuine field, is the setting in which the mechanism is measured at full strength.

\section{Baselines}
\label{appendix:baselines}

\begin{itemize}
  \item \textbf{DIHN}~\cite{10.1145/3485447.3511970} is an early model designed for the trigger-induced setting, which it casts as a CTR task. A dedicated module scores how strongly the user is currently committed to the trigger, and that score modulates both a trigger--target fusion and a trigger-aware aggregation of the behaviour history.

  \item \textbf{DIAN}~\cite{10.1145/3543873.3584661} turns the balance between trigger and history into an explicit decision. The network judges whether the request is driven by the trigger or by longer-term preference, and the trigger-based and trigger-free predictions are then merged through a probability decomposition over these two cases rather than through a learned gate.

  \item \textbf{DEI2N}~\cite{10.1145/3616855.3635829} argues that the instant interest revealed by the trigger does not stay constant. Alongside an intent estimator in the spirit of DIHN, it adds a time-aware component and an explicit trigger--target interaction unit, so that the weight given to the trigger can drift as the sequence progresses.

  \item \textbf{GRU4Rec}~\cite{hidasi2016sessionbasedrecommendationsrecurrentneural} encodes the user behaviour sequence with gated recurrent units and predicts the next item from the final hidden state. It is the standard recurrent baseline for session-based and sequential recommendation.

  \item \textbf{SASRec}~\cite{8594844} replaces recurrence with a causal self-attention stack, which lets every position attend to all earlier behaviours, and is trained with next-item prediction over the item ID vocabulary.

  \item \textbf{TIGER}~\cite{rajput2023recommender} is the representative Semantic ID generative recommender. Items are quantized into hierarchical discrete codes, and an encoder--decoder Transformer autoregressively generates the code sequence of the next item, which is then decoded back to an item by beam search over the code space.

  \item \textbf{HSTU}~\cite{3692070.3694484} formulates recommendation as sequential transduction over a unified stream of user actions, with an attention design tailored to long, non-stationary behaviour sequences.

  \item \textbf{LC-Rec}~\cite{10597986} adapts a large language model to Semantic ID recommendation. On top of the quantized indices it introduces a set of alignment tuning tasks that bridge the language space of the backbone and the newly introduced item index tokens, and then recommends by generating index tokens.

  \item \textbf{GR + trigger append} is a simple generative baseline that appends the trigger item as a single terminal token to the historical sequence within the standard encoder self-attention mechanism, evaluating whether directly exposing the trigger to the encoder is sufficient.

  \item \textbf{GR + trigger conditioning} keeps the historical sequence encoder unchanged and projects the trigger representation via a shared feed-forward layer, adding it directly to the token embeddings at every autoregressive decoding step to assess uniform condition steering.

  \item \textbf{GR + trigger-history attention} uses the trigger representation as a cross-attention query over the pristine historical sequence to construct an aggregated context vector for generation, verifying whether conventional attention pooling without hierarchical decoding adaptation is adequate.
\end{itemize}

\section{Semantic ID Construction}
\label{app:sid}

The title and description of each item are concatenated and encoded by a pretrained text embedding model, whose last hidden states are mean pooled into one vector $x_i \in \mathbb{R}^{2560}$. A residual-quantized autoencoder~\cite{9879532} then maps it to $z_i = \mathrm{Enc}(x_i) \in \mathbb{R}^{32}$ and quantizes it with $L = 3$ codebooks $E^{(l)} \in \mathbb{R}^{K \times 32}$ of size $K = 256$. Starting from $r_1 = z_i$, every level selects its nearest code and passes the residual on:
\begin{equation}
  c^{(l)}_i = \arg\min_{k} \bigl\| r_l - E^{(l)}_k \bigr\|_2 ,
  \qquad r_{l+1} = r_l - E^{(l)}_{c^{(l)}_i} ,
\end{equation}
so that item $i$ is represented by the tuple $(c^{(1)}_i, c^{(2)}_i, c^{(3)}_i)$ and reconstructed from $\hat{z}_i = \sum_l E^{(l)}_{c^{(l)}_i}$. The tokenizer is trained with a reconstruction term and the usual quantization terms,
\begin{equation}
  \begin{split}
    \mathcal{L} = {}& \bigl\| x_i - \mathrm{Dec}(\hat{z}_i) \bigr\|^2 \\
    &+ \sum_{l} \Bigl( \bigl\| \mathrm{sg}[r_l] - E^{(l)}_{c^{(l)}_i} \bigr\|^2
      + \beta \bigl\| r_l - \mathrm{sg}[E^{(l)}_{c^{(l)}_i}] \bigr\|^2 \Bigr) ,
  \end{split}
\end{equation}
with $\beta = 0.25$, $\mathrm{sg}[\cdot]$ the stop-gradient operator and codebooks initialized by $k$-means; a Sinkhorn-based assignment~\cite{2999792.2999868} on the last level spreads items that would otherwise share an identical tuple. Semantic IDs are produced once and then frozen: during recommender training they receive no gradient and act purely as a discrete input and supervision target.

Quantization is not injective, so a few items still collide: $3{,}686$ items map to $3{,}670$ distinct tuples on our data. We therefore keep a one-to-many index, and at evaluation time every generated tuple is expanded into all of its items, which occupy consecutive slots of the ranked list. A one-to-one mapping would make the losing items unreachable and understate Recall; discarding the groups would inflate it.

\section{Implementation Details}
\label{app:impl}

\begin{table}[!t]
  \caption{Hyper-parameters of TriGen on the public dataset.}
  \label{tab:hparam}
  \begin{tabular}{@{}l>{\raggedright\arraybackslash}p{0.63\columnwidth}@{}}
    \toprule
    Item & Value \\
    \midrule
    Semantic ID     & $3$ levels $\times$ $256$ codes, token dim.\ 128 \\
    hidden / FFN    & 128 / 320 \\
    layers          & 2 encoder, 2 decoder, 4 attention heads \\
    position enc.\  & RoPE, $\theta = 10^4$, dropout 0.1 \\
    history length  & 12, trigger at the tail \\
    output head     & cosine softmax, $\tau = 20$, tied weights \\
    fusion          & convex, router bias 2.0 \\
    parameters      & 1.51M \\
    \midrule
    optimizer       & AdamW, lr $10^{-3}$, weight decay $10^{-4}$ \\
    schedule        & 1000 warmup steps, cosine decay \\
    batch / epochs  & 256 / 15, gradient clipping 1.0 \\
    selection       & best validation Recall@20 \\
    inference       & beam search, size 20, trie-constrained \\
    \bottomrule
  \end{tabular}
\end{table}

Table~\ref{tab:hparam} lists the configuration used on the public dataset; all variants share it and differ only in the component under study. The public-data construction retains at most 10 history items; extracting the trigger leaves at most 9, which are left-padded to 12 positions. The encoder uses left padding, so the most recent behaviour sits at the right end of the sequence and the tail injection is adjacent to it in RoPE terms. Since one leftmost padding position is dropped for the trigger, the encoder length stays at $M = 12$ for every variant, so trigger-aware and trigger-free models never differ in sequence budget. On the industrial dataset, the trigger is repeated three times at the tail to strengthen its influence amid long histories containing many valid behavioural items; on the public dataset, the histories are short, so the trigger is inserted only once without repetition. Decoding is constrained by a trie of all valid Semantic IDs. On the industrial dataset, Semantic IDs consist of $8$ levels with $300$ codes each (vs.\ $3 \times 256$ on the public dataset). Note that the industrial dataset is evaluated under a multi-positive recall setting, as a single trigger-induced session may contain multiple positive target items. Every reported number is the mean over five seeds, with standard deviations reported in Appendix~\ref{sec:appendix_stds}. Public-dataset experiments run on a single NVIDIA A100 GPU with PyTorch 2.4.0 and CUDA 12.1; the industrial models are trained on eight NVIDIA L20 GPUs.

\begin{table*}[h!]
  \centering
  \caption{Standard deviations across 5 independent runs on the industrial dataset and the Amazon \textit{Industrial\_and\_Scientific} benchmark. The corresponding means are reported in Table~\ref{tab:main_results}. R@$K$ and N@$K$ denote Recall@$K$ and NDCG@$K$, respectively.}
  \label{tab:stds_results}
  \begin{tabularx}{\textwidth}{@{}ll*{8}{>{\centering\arraybackslash}X}@{}}
    \toprule
    \multirow{2}{*}{Category} & \multirow{2}{*}{Method} &
    \multicolumn{4}{c}{Industrial Dataset} &
    \multicolumn{4}{c}{Amazon \textit{Industrial\_and\_Scientific}} \\
    \cmidrule(lr){3-6} \cmidrule(lr){7-10}
    & & R@5 & R@10 & N@5 & N@10 & R@5 & R@10 & N@5 & N@10 \\
    \midrule
    \multirow{3}{*}{TIR-specific}
      & DIHN                          & 0.01 & 0.02 & 0.01 & 0.02 & 0.18 & 0.24 & 0.18 & 0.21 \\
      & DIAN                          & 0.01 & 0.02 & 0.01 & 0.02 & 0.11 & 0.34 & 0.16 & 0.13 \\
      & DEI2N                         & 0.02 & 0.03 & 0.01 & 0.02 & 0.14 & 0.25 & 0.08 & 0.11 \\
    \midrule
    \multirow{2}{*}{Sequential}
      & GRU4Rec                       & 0.01 & 0.02 & 0.01 & 0.02 & 0.16 & 0.19 & 0.13 & 0.12 \\
      & SASRec                        & 0.01 & 0.02 & 0.01 & 0.02 & 0.18 & 0.22 & 0.14 & 0.14 \\
    \midrule
    \multirow{3}{*}{\shortstack[l]{Generative\\(Trigger-free)}}
      & HSTU                          & 0.02 & 0.02 & 0.01 & 0.02 & 0.13 & 0.18 & 0.12 & 0.14 \\
      & TIGER                         & 0.02 & 0.02 & 0.01 & 0.02 & 0.21 & 0.38 & 0.19 & 0.18 \\
      & LC-Rec                        & 0.01 & 0.02 & 0.01 & 0.02 & 0.29 & 0.15 & 0.13 & 0.09 \\
    \midrule
    \multirow{3}{*}{Simple TIR-GR}
      & GR + trigger append           & 0.02 & 0.03 & 0.01 & 0.02 & 0.35 & 0.26 & 0.22 & 0.16 \\
      & GR + trigger conditioning     & 0.02 & 0.03 & 0.01 & 0.02 & 0.17 & 0.08 & 0.16 & 0.12 \\
      & GR + trigger-history attention& 0.02 & 0.03 & 0.02 & 0.03 & 0.39 & 0.35 & 0.32 & 0.28 \\
    \midrule
    Ours
      & TriGen               & 0.02 & 0.03 & 0.01 & 0.02 & 0.24 & 0.31 & 0.18 & 0.15 \\
    \bottomrule
  \end{tabularx}
\end{table*}

\section{Statistical Significance and Standard Deviations}
\label{sec:appendix_stds}

To assess reproducibility and statistical stability, all models are trained and evaluated across five independent runs with distinct random seeds. The main text reports the mean metrics across these runs in Table~\ref{tab:main_results}. Table~\ref{tab:stds_results} reports the corresponding standard deviations on both the industrial dataset and the Amazon benchmark.

Given that the industrial evaluation is conducted over millions of user interactions, the empirical variances remain marginal (within $\pm 0.01$ to $\pm 0.03$), providing strong statistical confidence in the measured improvements. In contrast, standard deviations on the public dataset exhibit wider ranges due to its smaller sample size, which aligns with expected statistical sampling behavior. Across all runs, TriGen sustains consistent gains over both discriminative and generative baselines.

\section{Limitations}
\label{app:limit}

\textbf{The public trigger is simulated.} No public dataset with Semantic-ID-ready item text exposes a real trigger field, so on Amazon the trigger is the last history item rather than an observed click. Baselines keep the full sequence, so this does not favour TriGen; it instead caps the trigger signal available to it, and the public gap should be read as conservative. Only the industrial dataset, where the trigger is an independent field, measures the mechanism at full strength.

\textbf{Only a single trigger is modelled.} TriGen assumes exactly one trigger per request, which matches the detail-page scenario but not surfaces where several recent signals arrive at once. Extending both injection points to a set of triggers, and how the fusion arbitrates among them, is left to future work.

\end{document}